\documentclass[showpacs,preprintnumbers,amsmath,amssymb,floatfix]{revtex4-2}
\usepackage{graphicx}
\usepackage{dcolumn}
\usepackage{bm}
\begin{document}
\title{Decoupling of the structure functions in momentum space based on the Laplace transformation }

\author{G.R.Boroun}%
 \email{boroun@razi.ac.ir }
 \affiliation{Department of Physics, Razi University, Kermanshah
67149, Iran}%
\author{Phuoc Ha }
\altaffiliation{pdha@towson.edu}
\affiliation{Department of Physics, Astronomy and Geosciences, Towson University, Towson, MD 21252}
\date{\today}
\begin{abstract}
Using Laplace transform techniques, we describe the determination of the longitudinal structure function
$F_{L}(x,Q^2)$, at the leading-order approximation in momentum space, from the structure function $F_{2}(x,Q^2)$ and its derivative with respect to ${\ln}Q^2$
in a kinematical region of low values of the Bjorken variable $x$.
Since the $x$ dependence of $F_2(x,Q^2)$ and its evolution with $Q^2$ are determined much better by the data than $F_L(x,Q^2)$, this
method provides both a direct check on $F_L(x,Q^2)$ where measured, and a way of extending $F_L(x,Q^2)$ into regions of $x$ and $Q^2$ where there are currently no data.
In our calculations, we ultilize the Block-Durand-Ha parametrization
for the structure function $F_{2}(x,Q^2)$  [M. M. Block, L. Durand and P. Ha, Phys.Rev.D {\bf89}, 094027 (2014)].
We find that the Laplace transform method in momentum space provides correct
behaviors of the extracted longitudinal structure function $F_{L}(x,Q^2)$ and that
our obtained results are in line with data from the H1 Collaboration
and other results for $F_{L}(x,Q^2)$ obtained using Mellin transform method.

\end{abstract}
\maketitle
\section{Introduction}
Recently,  evolution of the longitudinal and transversal
structure functions in momentum space has been considered in \cite{Ref1}.
Structure functions measurable in deep inelastic scattering (DIS)
are formulated in the momentum-space
Dokshitzer-Gribov-Lipatov-Altarelli-Parisi (DGLAP) evolution
equations \cite{Ref2, Ref3, Ref4}. Scheme independent evolution equations for the
structure functions $ F_{i}(x,Q^{2})$ proposed some time ago in \cite{Ref5, Ref6}, as
the physical observables, read
\begin{eqnarray}
\frac{{\partial}F_{i}(x,Q^{2})}{{\partial}{\ln}Q^{2}}=
\sum_{j}P_{ij}\otimes F_{j}(x,Q^{2})
\end{eqnarray}
where anomalous dimensions, $P_{ij}$, are computable in
perturbative QCD (pQCD). Determination of the longitudinal
structure function in the nucleon from the proton structure
function, based on a form of the deep inelastic lepton-hadron
scattering  structure function proposed by
Block-Durand-Ha (BDH) in \cite{Ref7}, is considered in \cite{ Ref8, Ref9, Ref10, Ref11}.
Parametrization of the proton structure function proposed in \cite{Ref7} describes the available experimental data
on the reduced cross sections at low $x$ and provides a behavior
of the hadron-hadron cross sections ${\sim}{\ln^{2}}s$ at large
$s$ in a full accordance with the Froissart predictions \cite{Ref12} ($s$
is the Mandelstam variable denoting the square of the total
invariant energy of the process).

Deep inelastic scattering (DIS) is characterized by structure
functions $F_{k}$ that depend on kinematic variables Bjorken $x$
and momentum transfer $Q$ by the following form
\begin{eqnarray}
F_{k}(x,Q^2)=<e^2>\sum_{a=s,g}\bigg{[}C_{k,a}(x,\alpha_{s}){\otimes}xf_{a}(x,Q^2)\bigg{]},~k=2,L~
\end{eqnarray}
where $C_{k,a}(x,\alpha_{s})$ are the known Wilson coefficient
functions in the order of the perturbation theory, $\alpha_{s}$
is the strong coupling, and $<e^{2}>$ is
the average of the charge $e^{2}$ for the active quark flavors,
$<e^{2}>=n_{f}^{-1}\sum_{i=1}^{n_{f}}e_{i}^{2}$ with $n_{f}$ as
the number of considered flavors. The symbol $\otimes$ denotes
convolution according to the usual prescription and
$xf_{a=2,g}(x,Q^2)$ are the singlet-quark and gluon densities
respectively (the non-singlet quark distributions at small $x$
become negligibly small in comparison with the singlet
distributions). The DGLAP equations, which describe how the parton
distribution functions (PDFs) vary as the energy scale of the
scattering process changes, are important for understanding a wide
range of high-energy processes, including DIS, hadron collisions,
and deep-inelastic scattering of heavy ions at future colliders
(Large Hadron electron Collider (LHeC) \cite{Ref13} and Electron-Ion
Collider (EIC) \cite{Ref14, Ref15}).

Numerical and analytical methods (which extract the PDFs from the
experimental data) to solve the DGLAP evolution equations  have
been extensively studied in the literature \cite{Ref16, Ref17, Ref18,
Ref19, Ref20, Ref21, Ref22, Ref23, Ref24, Ref25, Ref26, Ref27, Ref28,
Ref29, Ref30, Ref31, Ref32}. The solutions
to these equations provide a theoretical prediction for the PDFs
used in the interpretation and description of the Hadron-Electron
Ring Accelerator (HERA) data on the total and diffractive cross-sections in deep inelastic electron-proton
scattering. They serve as a mean to test our understanding of QCD and extract
information about the structure of the proton.

In this paper, we extend the method using a Laplace-transform
technique and obtain an analytical method for the solution of the
momentum-space of the DGLAP equations for $F_{L}(x,Q^2)$ in terms
of $F_{2}(x,Q^2)$ and known derivative $dF_{2}(x,Q^2)/d{\ln}Q^2$
in the kinematical region of low values of the Bjorken variable
$x$. The parameterization of the structure function $F_{2}(x,Q^2)$
in \cite{Ref7} is obtained from a combined fit to HERA data in a wide
range of the kinematical variables $x$ and $Q^2$ by the following
explicit expression as
\begin{eqnarray}
F^{\gamma p}_{ 2}(x,Q^{2})=D(Q^{2})(1-
x)^{n}\bigg{[}C(Q^{2})+A(Q^{2})\ln(\frac{1}{x}\frac{Q^{2}}{Q^{2}+\mu^{2}})+B(Q^{2})\ln^{2}(\frac{1}{x}\frac{Q^{2}}{Q^{2}+\mu^{2}})\bigg{]},
\end{eqnarray}
where
\begin{eqnarray}
 A(Q^{2})& =& a_{0} + a_{1} {\ln}(1+\frac{Q^{2}}{\mu^{2}}) + a_{2} {\ln}^{2}(1+\frac{Q^{2}}{\mu^{2}})
 ,\nonumber\\
B(Q^{2})& =& b_{0} + b_{1} {\ln}(1+\frac{Q^{2}}{\mu^{2}}) +
b_{2} {\ln}^{2}(1+\frac{Q^{2}}{\mu^{2}})
 ,\nonumber\\
C(Q^{2})& =& c_{0} + c_{1}
 {\ln}(1+\frac{Q^{2}}{\mu^{2}}),\nonumber\\
D(Q^{2})& =& \frac{Q^{2}(Q^{2}+\lambda M^{2})}{(Q^{2}+M^{2})^2},
\end{eqnarray}
where $M$ is the effective mass and $\mu^{2}$ is a scale factor
defined by the Block-Halzen fit to the real photon-proton cross
section \cite{Ref33} in Table I. In the following, we apply this
parametrization function to test the consistency of the
longitudinal structure function owing to the momentum-space of the
DGLAP equations with HERA data on deep inelastic electron-proton
scattering.

The paper is organized as follows: in section II, we present the
basics of the momentum-space of the DGLAP equations. Section III
summarizes the Laplace transform method for obtaining an analytical
solution for the longitudinal structure function. In section IV,
the numerical results are obtained and compared with the available
H1 Collaboration data \cite{Ref34, Ref35, Ref36} and the Large Hadron electron
Collider (LHeC) \cite{Ref13} simulated errors.
Conclusions are given in Sec. V.  Some detailed calculations are relegated to Appendix.

\section{Momentum Space}

The DIS structure functions $F_{2}$ and $F_{L}$ at low $x$ are
defined \cite{Ref1} into the singlet and gluon distribution
functions by the following forms
\begin{eqnarray}
F_{2}(x,Q^2)&=&<e^2>\bigg{\{}
C^{(0)}_{2,s}+\frac{\alpha_{s}(\mu_{r}^{2})}{2\pi}\bigg{[}C^{(1)}_{2,s}
-\ln{\bigg{(}}\frac{\mu_{r}^{2}}{Q^2}{\bigg{)}}C^{(0)}_{2,s}{\otimes}P_{qq}
\bigg{]}\bigg{\}}{\otimes}x\Sigma(x,\mu_{r}^{2})\nonumber\\
&&+2\sum_{i=1}^{n_{f}}e_{i}^{2}\frac{\alpha_{s}(\mu_{r}^{2})}{2\pi}\bigg{[}C^{(1)}_{2,g}
-\ln{\bigg{(}}\frac{\mu_{r}^{2}}{Q^2}{\bigg{)}}C^{(0)}_{2,g}{\otimes}P_{qg}
\bigg{]}{\otimes}xg(x,\mu_{r}^{2}),
\end{eqnarray}
and
\begin{eqnarray}
F_{L}(x,Q^2)&=&<e^2>\frac{\alpha_{s}(\mu_{r}^{2})}{2\pi}\bigg{\{}
C^{(1)}_{L,s}+\frac{\alpha_{s}(\mu_{r}^{2})}{2\pi}\bigg{[}C^{(2)}_{L,s}
-\ln{\bigg{(}}\frac{\mu_{r}^{2}}{Q^2}{\bigg{)}}C^{(1)}_{L,s}{\otimes}P_{qq}
-2n_{f}\ln{\bigg{(}}\frac{\mu_{r}^{2}}{Q^2}{\bigg{)}}C^{(1)}_{L,g}{\otimes}P_{gq}\bigg{]}\bigg{\}}{\otimes}x\Sigma(x,\mu_{r}^{2})\nonumber\\
&&+2\sum_{i=1}^{n_{f}}e_{i}^{2}\frac{\alpha_{s}(\mu_{r}^{2})}{2\pi}\bigg{\{}C^{(1)}_{L,g}+
\frac{\alpha_{s}(\mu_{r}^{2})}{2\pi}\bigg{[}C^{(2)}_{L,g}-\ln{\bigg{(}}\frac{\mu_{r}^{2}}{Q^2}{\bigg{)}}C^{(1)}_{L,s}{\otimes}P_{qg}
-\ln{\bigg{(}}\frac{\mu_{r}^{2}}{Q^2}{\bigg{)}}C^{(1)}_{L,g}{\otimes}P_{gg}
\bigg{]}\bigg{\}}{\otimes}xg(x,\mu_{r}^{2})\nonumber\\
&&+<e^2>\bigg{(}\frac{\alpha_{s}(\mu_{r}^{2})}{2\pi}\bigg{)}^2\bigg{[}b_{0}\ln{\bigg{(}}\frac{\mu_{r}^{2}}{Q^2}{\bigg{)}}
\bigg{]}\bigg{[}C^{(1)}_{L,s}{\otimes}x\Sigma(x,\mu_{r}^{2})+
2n_{f}C^{(1)}_{L,g}{\otimes}xg(x,\mu_{r}^{2}) \bigg{]},
\end{eqnarray}
where $xg(x,\mu_{r}^{2})$ and
$x\Sigma(x,\mu_{r}^{2}){\equiv}xf_{s}(x,\mu_{r}^{2})=\sum_{q}[xq(x,\mu_{r}^{2})+x\overline{q}(x,\mu_{r}^{2})]$
are the gluon and singlet distribution functions at the
renormalization scale $\mu_{r}^{2}$, respectively. In Eqs.(5) and
(6), $C_{ij} (i=2,L; j=s,g)$ denote the scheme-dependent coefficient functions
defined, at the first non-zero
order in $\alpha_{s}$, by [1]
\begin{eqnarray}
C^{(0)}_{2,s}(x)&=&\delta(1-x),\nonumber\\
C^{(1)}_{L,s}(x)&=&2C_{F}x,\nonumber\\
C^{(1)}_{L,g}(x)&=&4T_{R}x(1-x),
\end{eqnarray}
with the color factors $C_{A}=3$, $T_{R}=1/2$ and $C_{F}=4/3$
associated with the color group SU(3). The authors in \cite{Ref1}
inverted the leading non-zero order part of Eqs.(5) and (6). As a
result, the singlet and gluon densities in terms of the structure
functions read
\begin{eqnarray}
\Sigma(x,\mu_{r}^{2})&=&\frac{1}{<e^2>}\frac{F_{2}(x,Q^2)}{x},\nonumber\\
g(x,\mu_{r}^{2})&=&\frac{1}{\sum_{i=1}^{n_{f}}e_{i}^{2}}\bigg{(}\frac{C_{F}}{4T_{R}}\delta(1-x){\otimes}x\frac{d}{dx}\frac{F_{2}(x,Q^2)}{x}
-\frac{C_{F}}{2T_{R}}\delta(1-x){\otimes}\frac{F_{2}(x,Q^2)}{x} +
\frac{1}{8T_{R}}\frac{2\pi}{\alpha_{s}(\mu_{r}^{2})}\delta(1-x){\otimes}x^2\frac{d^2}{dx^2}\frac{F_{L}(x,Q^2)}{x}\nonumber\\
&&-\frac{1}{4T_{R}}\frac{2\pi}{\alpha_{s}(\mu_{r}^{2})}\delta(1-x){\otimes}x\frac{d}{dx}\frac{F_{L}(x,Q^2)}{x}
+\frac{1}{4T_{R}}\frac{2\pi}{\alpha_{s}(\mu_{r}^{2})}\delta(1-x){\otimes}\frac{F_{L}(x,Q^2)}{x}\bigg{)}.
\end{eqnarray}
 Using the leading-order (LO) renormalization group equation
 \begin{eqnarray}
\mu_{r}^2\frac{d\alpha_{s}(\mu_{r}^{2})}{d\mu_{r}^2}=-(\frac{11C_{A}-4T_{R}n_{f}}{12\pi})\alpha^{2}_{s}(\mu_{r}^{2}) ,
 \end{eqnarray}
 and setting the renormalization scale equal to the momentum
 transfer, $\mu_{r}^{2}=Q^2$, the authors in [1] derived the evolution equation of
the structure function $F_{2}(x,Q^2)$ at
 the first non-zero order in $\alpha_{s}$ as
\begin{eqnarray}
\frac{dF_{2}(x,Q^2)}{d{\ln}Q^2}&=&\frac{\alpha_{s}(Q^2)}{2\pi}x\bigg{\{}
\frac{1}{4}\bigg{(}\frac{2}{x}-\frac{d}{dx}\bigg{)}\frac{2\pi}{\alpha_{s}(Q^2)}F_{L}(x,Q^2)
+\frac{1}{2}\int_{x}^{1}\frac{dz}{z^2}\frac{2\pi}{\alpha_{s}(Q^2)}F_{L}(z,Q^2)\nonumber\\
&&+C_{F}\bigg{[}\frac{1}{x}F_{2}(x,Q^2)-\int_{x}^{1}\frac{dz}{z^2}\bigg{(}1-\frac{x}{z}\bigg{)}F_{2}(z,Q^2)
+\frac{1}{x}\int_{x}^{1}dz\frac{1+z^2}{(1-z)_{+}}F_{2}(\frac{x}{z},Q^2)\bigg{]}\bigg{\}},
\end{eqnarray}
where the plus function is defined as
\begin{eqnarray}
\int_{x}^{1}dz\frac{f(z)}{(1-z)_{+}}=\int_{x}^{1}dz \frac{f(z)-f(1)}{1-z}+f(1)\ln(1-x).
\end{eqnarray}
By writing
\begin{equation}
\frac{1+z^2}{(1-z)_{+}}=-(1+z)+\frac{2}{(1-z)_{+}} ,
\end{equation}
we can rewrite the last integral in Eq. (10) as
\begin{eqnarray}
\frac{1}{x}\int_{x}^{1}dz\frac{1+z^2}{(1-z)_{+}}F_{2}(\frac{x}{z},Q^2) &=&-\frac{1}{x}\int_{x}^{1}dz(1+z)F_{2}(\frac{x}{z},Q^2)+\frac{1}{x}\int_{x}^{1}dz\frac{2}{(1-z)_{+}}F_{2}(\frac{x}{z},Q^2) )\nonumber\\
&=&-\int_{x}^{1}\frac{dz}{z^2}\bigg{(}1+\frac{x}{z}\bigg{)}F_{2}(z,Q^2)+\frac{1}{x}\int_{x}^{1}dz\frac{2}{(1-z)_{+}}F_{2}(\frac{x}{z},Q^2) ).
\end{eqnarray}
Substituting Eq. (13) into Eq. (10), we get
\begin{eqnarray}
\frac{dF_{2}(x,Q^2)}{d{\ln}Q^2}&=&\frac{\alpha_{s}(Q^2)}{2\pi}x\bigg{\{}
\frac{1}{4}\bigg{(}\frac{2}{x}-\frac{d}{dx}\bigg{)}\frac{2\pi}{\alpha_{s}(Q^2)}F_{L}(x,Q^2)
+\frac{1}{2}\int_{x}^{1}\frac{dz}{z^2}\frac{2\pi}{\alpha_{s}(Q^2)}F_{L}(z,Q^2)\nonumber\\
&&+C_{F}\bigg{[}\frac{1}{x}F_{2}(x,Q^2)-2\int_{x}^{1}\frac{dz}{z^2}F_{2}(z,Q^2)
+\frac{2}{x}\int_{x}^{1}dz\frac{1}{(1-z)_{+}}F_{2}(\frac{x}{z},Q^2)\bigg{]}\bigg{\}},
\end{eqnarray}

\section{ Laplace Transformation}

In the following, we use the method developed in detail in
\cite{Ref37, Ref38, Ref39, Ref40} to obtain the longitudinal
structure function into the proton structure function and its
derivative using a Laplace-transform method. We now rewrite the
momentum-space DGLAP evolution equation for the longitudinal
structure function (i.e., Eq.(14)) in terms of the variables
$\upsilon={\ln}(1/x)$ and $Q^2$ instead of $x$ and $Q^2$. Using
the notation
$\widehat{F_{i}}(\upsilon,Q^2){\equiv}F_{i}(e^{-\upsilon},Q^2)$
for structure functions, explicitly, from Eq.(14), we find
\begin{eqnarray}
\frac{d\widehat{F}_{2}(\upsilon,Q^2)}{d{\ln}Q^2}&=&\bigg{(}\frac{1}{2}+\frac{1}{4}\frac{d}{d\upsilon}\bigg{)}\widehat{F}_{L}(\upsilon,Q^2)
+\frac{1}{2}\int_{0}^{\upsilon}{dw}~e^{-(\upsilon-w)}\widehat{F}_{L}(w,Q^2)+C_{F}\frac{\alpha_{s}(Q^2)}{2\pi}\widehat{F}_{2}(\upsilon,Q^2)\nonumber\\
&&-2C_{F}\frac{\alpha_{s}(Q^2)}{2\pi}\int_{0}^{\upsilon}{dw}~e^{-(\upsilon-w)}\widehat{F}_{2}(w,Q^2)+2C_{F}\frac{\alpha_{s}(Q^2)}{2\pi}
\int_0^v dw \ln\left(1-e^{-(v-w)}\right)\frac{\partial
\widehat{F}_{2}(w,Q^2)}{\partial w} .
\end{eqnarray}
Introducing the notation that the Laplace transform of structure
functions $\widehat{F}_{i}(\upsilon,Q^2)$ are given by
${f}_{i}(s,Q^2)$ as
${f}_{i}(s,Q^2){\equiv}{\mathcal{L}}[\widehat{F}_{i}(\upsilon,Q^2);s]$
and using the fact that the Laplace transform of a convolution
factors is simply the ordinary product of the Laplace transform of
the factors by the following form
\begin{eqnarray}
{\mathcal{L}}\bigg{[}\int_{0}^{\upsilon}\widehat{F}_{i}(w,Q^2)\widehat{H}(\upsilon-w)
dw;s\bigg{]}={f}_{i}(s,Q^2){\times}h(s),
\end{eqnarray}
where
$h(s){\equiv}{\mathcal{L}}[e^{-\upsilon}\widehat{H}(\upsilon)]$,
we find that the Laplace transform of Eq. (15) is given by
\begin{eqnarray}
\frac{d{F}_{2}(s,Q^2)}{d{\ln}Q^2}&=&\bigg{(}\frac{1}{2}+\frac{1}{4}s+\frac{1}{2}\frac{1}{1+s}\bigg{)}F_{L}(s,Q^2)-\frac{1}{4}F_{L}(0,Q^2)\nonumber\\
&&+C_{F}\frac{\alpha_{s}(Q^2)}{2\pi}\bigg{(}1-\frac{2}{1+s} -2
H_s\bigg{)}F_{2}(s,Q^2)+2C_{F}\frac{\alpha_{s}(Q^2)}{2\pi}\frac{H_{s}}{s}F_{2}(0,Q^2),
\end{eqnarray}
where  $H_s=\psi(s+1)-\psi(1)$ with
$\psi(s)=\Gamma'(s)/\Gamma(s)$. Here
$\psi(1)=-\gamma=-0.5772156...$ is Euler constant. The functions
$F_{i}(0)$ (or exactly $F_{i}(+0)$) are the boundary conditions
due to the Laplace derivatives and defined
$F_{i}(\upsilon=0,Q^2){\equiv}F_{i}(x=1,Q^2)=0$.
Then, from Eq. (17), we find
\begin{eqnarray}
F_{L}(s,Q^2)&=&h_{1}(s)\frac{d{F}_{2}(s,Q^2)}{d{\ln}Q^2}+{h_{2}(s)}F_{2}(s,Q^2),
\end{eqnarray}
where $h_1(s)= h_{L}^{-1}(s)$ and $h_2(s)=-C_{F}\frac{\alpha_{s}(Q^2)}{2\pi}\tilde h_{2}(s) h_L^{-1}(s)$
with  $h_{L}(s) $ and $\tilde h_{2}(s)$ given by
\begin{eqnarray}
h_{L}(s)&=&\frac{2+s}{4}+\frac{1}{2(1+s)},\nonumber\\
\tilde h_{2}(s)&=&1-\frac{2}{1+s}-2H_s.
\end{eqnarray}
The inverse Laplace transform of the coefficients $h_{i}$ in Eq.
(18), defined by the kernels
$\widehat{J}_{i}(\upsilon){\equiv}{\mathcal{L}^{-1}}[h_{i}(s);\upsilon]$,
is straightforward. We find that
\begin{eqnarray}
\widehat{J}_{1}(\upsilon)&=&4e^{-\frac{3}{2}\upsilon}\bigg{[}\cos{\bigg{(}}\frac{\sqrt{7}}{2}\upsilon{\bigg{)}}-\frac{\sqrt{7}}{7}
\sin{\bigg{(}}\frac{\sqrt{7}}{2}\upsilon{\bigg{)}}\bigg{]},
\end{eqnarray}
and
\begin{eqnarray}
\widehat{J}_{2}(\upsilon)&=&-4C_{F}\frac{\alpha_{s}(Q^2)}{2\pi}e^{-\frac{3}{2}\upsilon}\bigg{[}(1+2\psi(1))\cos{\bigg{(}}\frac{\sqrt{7}}{2}\upsilon{\bigg{)}}
-(5+2\psi(1))\frac{\sqrt{7}}{7}
\sin{\bigg{(}}\frac{\sqrt{7}}{2}\upsilon{\bigg{)}}\bigg{]}\nonumber\\
&&+8C_{F}\frac{\alpha_{s}(Q^2)}{2\pi} \widehat{f}(\upsilon) ,
\end{eqnarray}
where
\begin{eqnarray}
\widehat{f}(\upsilon) &=& {\mathcal{L}^{-1}}\bigg{[}\frac{(s+1)\psi(s+1)}{4+3s+s^2},s;\upsilon\bigg{]} \nonumber\\
&& =- 2 e^{-\frac{3}{2}  \upsilon} \bigg{[} 0.1704 \cos
\bigg{(}\frac{\sqrt{7}}{2}\upsilon\bigg{)} +  1.211
\sin\bigg{(}\frac{\sqrt{7}}{2}\upsilon\bigg{)} \bigg{]} +
\sum_{k=1}^{\infty}\frac{k}{(k+1)^2-3(k+1)+4}  e^{-(k+1)
\upsilon} ,
\end{eqnarray}
as shown in the Appendix.
Transforming back in to $x$ space,  the longitudinal
structure function $F_{L}(x,Q^2)$ is given by
\begin{eqnarray}
F_{L}(x,Q^2)&=&4\int_{x}^{1}\frac{d{F}_{2}(z,Q^2)}{d{\ln}Q^2}(\frac{x}{z})^{3/2}\bigg{[}\cos{\bigg{(}}\frac{\sqrt{7}}{2}{\ln}\frac{z}{x}{\bigg{)}}-\frac{\sqrt{7}}{7}
\sin{\bigg{(}}\frac{\sqrt{7}}{2}{\ln}\frac{z}{x}{\bigg{)}}\bigg{]}\frac{dz}{z}\nonumber\\
&&-4C_{F}\frac{\alpha_{s}(Q^2)}{2\pi}\int_{x}^{1}F_{2}(z,Q^2)(\frac{x}{z})^{3/2}\bigg{[}(1.6817+2\psi(1))\cos{\bigg{(}}\frac{\sqrt{7}}{2}{\ln}\frac{z}{x}{\bigg{)}}
+(2.9542-2\frac{\sqrt{7}}{7}\psi(1))
\sin{\bigg{(}}\frac{\sqrt{7}}{2}{\ln}\frac{z}{x}{\bigg{)}}\bigg{]}\frac{dz}{z}\nonumber\\
&&+8C_{F}\frac{\alpha_{s}(Q^2)}{2\pi} \sum_{k=1}^{\infty}\frac{k}{(k+1)^2-3(k+1)+4}\int_{x}^{1}F_{2}(z,Q^2)(\frac{x}{z})^{k+1} \frac{dz}{z}.
\end{eqnarray}

\section{Results and Discussions}

With the explicit form of the longitudinal structure function
(i.e., Eq. (23)), we begin to extract the numerical results at
small $x$ in a wide range of $Q^2$ values, using the
parametrization of $F_{2}(x,Q^2)$ given by Eq. (3). The QCD
parameter $\Lambda$ for four numbers of active flavor has been
extracted \cite{Ref10} due to $\alpha_{s}(M_{z}^{2})=0.1166$ with
respect to the LO form of $\alpha_{s}(Q^2)$ with
$\Lambda=136.8~\mathrm{MeV}$. In order to make the effect of
production threshold for charm quark
 with $m_{c}=1.29^{+0.077}_{-0.053}~\mathrm{GeV}$ \cite{Ref42, Ref43}, the
rescaling variable $\chi$ is defined by the form
$\chi=x(1+4\frac{m_{c}^{2}}{Q^2})$ where reduced to
the Bjorken variable $x$ at high $Q^2$ \cite{Ref43}.

\begin{figure}[h]
\includegraphics[width=0.6\textwidth]{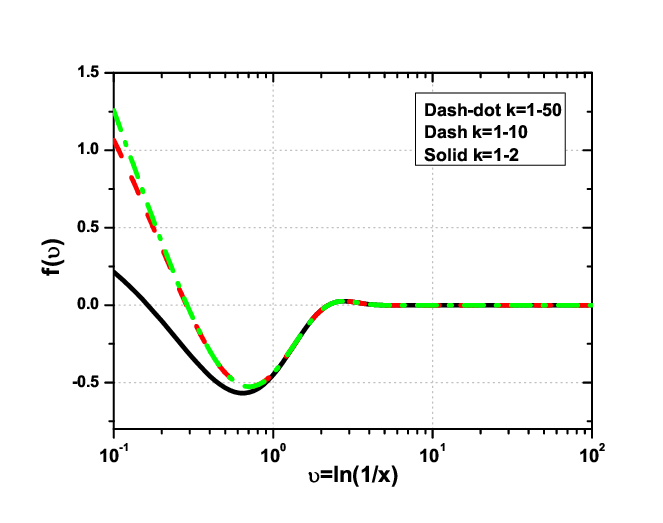}
\caption{The function $\widehat{f}(\upsilon)$ [i.e., Eq.(22)] is
plotted in a wide range of $\upsilon$. The solid black, dashed
red, and dot-dashed green curves are  shown according to the
expansion of the second term in Eq.(22) in the ranges $k=1-2$,
$k=1-10$, and $k=1-50$, respectively. The function
$\widehat{f}(\upsilon)$ diverges as ln $\upsilon$ for $\upsilon
\to 0$.}\label{Fig1}
\end{figure}

In Fig. 1, we analyze the function $\widehat{f}(\upsilon)$ in a
wide range of $\upsilon$ according to the expansion of the second
term in Eq. (22) in the ranges $k=1-2$, $k=1-10$, and $k=1-50$,
respectively. We observe that the function $\widehat{f}(\upsilon)$
is very small for  $\upsilon \ge 2$ as would be expected from the
decreasing exponential factor in $\upsilon$ in Eq. (22). It is
nearly independent of the cutoff in the expansion for $\upsilon
\ge 1$, but the expansion must be carried to large $k$ for
$\upsilon$  small.
In the following we choose the value of $k=50$ in the
numerical results.
As seen in the figure, the result for  $\widehat{f}(\upsilon)$ appears
to converge well even for small $\upsilon > 0$ for the maximum value
of $ k$ sufficiently large. We have found that the choice $k=50$ for
the upper limit gives results sufficient accurate for our purposes.

In Fig. 2,  using the parametrization for $F_{2}(x,Q^2)$ in Eq.
(3), we have plotted the longitudinal structure function
$F_{L}(x,Q^2)$ from Eq. (23) as a function of $x$
($5{\times}10^{-5}<x<10^{-2}$) for the values of $Q^2=5, 15, 25$,
and $45~\mathrm{GeV}^2$, respectively. In the figure, the  H1
Collaboration data (H1 2011 \cite{Ref36} and H1 2014 \cite{Ref42})
accompanied by total errors for $Q^2=5, 15, 25$, and
$45~\mathrm{GeV}^2$ are also shown.

In Fig. 3, we have separated our analysis of the longitudinal
structure function at any fixed $Q^2$ and compare it with the
results in \cite{Ref8} and \cite{Ref10} at the LO approximation as
a function of $x$. The longitudinal structure function extracted
at $Q^2$ values are in good agreement with experimental data in
comparison with those in \cite{Ref8} at the LO approximation, as
the mathematical structure of Eq. (10) in momentum space differs
from the DGLAP equations for structure functions.

\begin{figure}[h]
\includegraphics[width=0.6\textwidth]{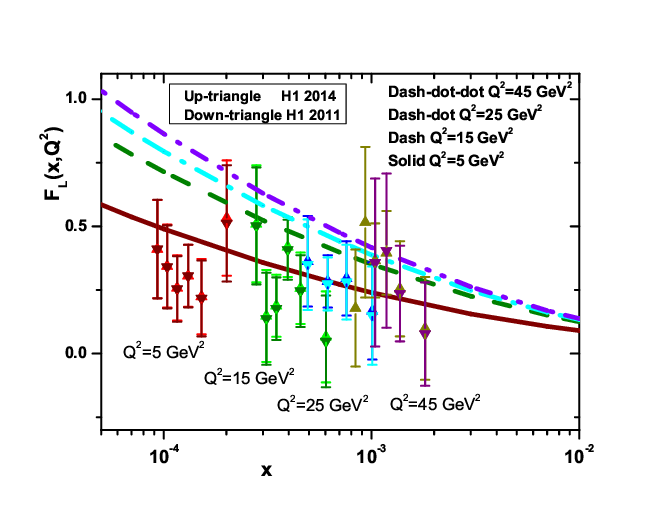}
\caption{Plots of the longitudinal structure function
$F_{L}(x,Q^2)$ versus $x$ for $Q^2=5$ (solid brown), $15$ (dash green), $25$ (dash-dot turquoise), and
$45~\mathrm{GeV}^2$ (dash-dot-dot purple), respectively. H1 Collaboration data are selected from
 \cite{Ref36}(H1 2011) and \cite{Ref42}(H1 2014). }\label{Fig2}
\end{figure}
\begin{figure}[h]
\includegraphics[width=0.7\textwidth]{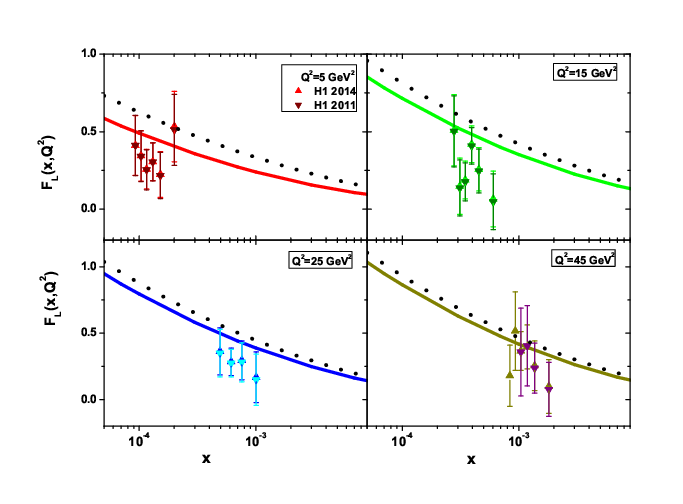}
\caption{The longitudinal structure function $F_{L}(x,Q^{2})$
(solid curves) plotted at fixed $Q^{2}$ as a function of $x$
variable, compared with the Mellin transforms method [8] (dot
curves) at the LO approximation. Experimental data (up-triangle H1
2014, down-triangle H1 2011) are from the H1-Collaboration \cite{Ref42, Ref36}
as accompanied with total errors. }\label{Fig3}
\end{figure}
\begin{figure}[h]
\includegraphics[width=0.7\textwidth]{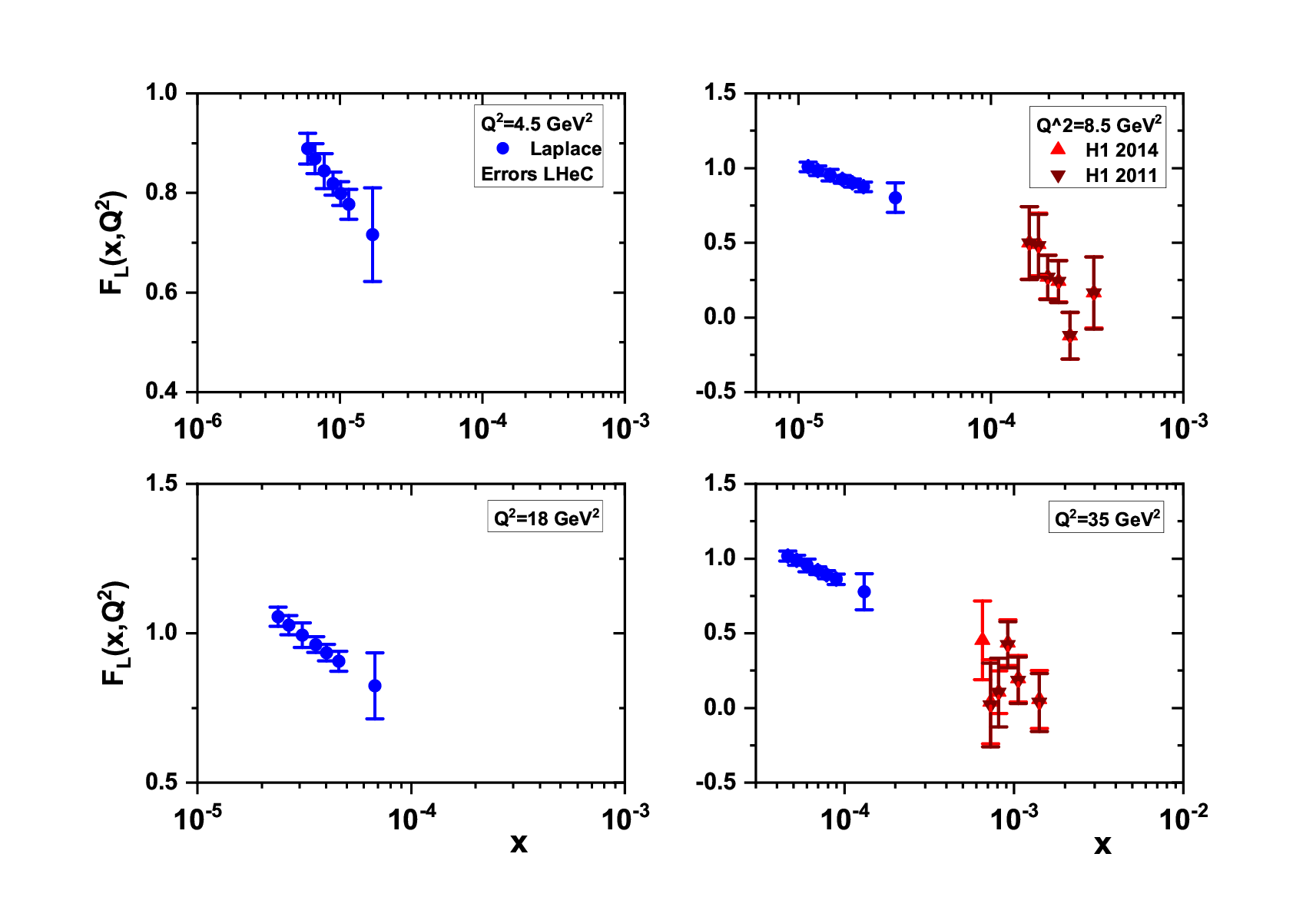}
\caption{The central values of the longitudinal structure function
are plotted at low $x$ as accompanied by the LHeC total errors
\cite{Ref13}. H1 Collaboration data \cite{Ref36,Ref42} are collected at $Q^2=8.5$ and
$35~\mathrm{GeV}^2$  with total errors. }\label{Fig4}
\end{figure}

In Fig. 4, the longitudinal structure functions in momentum space
at selected $x$ and $Q^2$ are associated with the LHeC simulated
uncertainties \cite{Ref13}. We observe that the longitudinal
structure functions (central values) are determined owing to Eq.
(23) for the $Q^{2}$ values (4.5, 8.5, 18, and
35$~\mathrm{GeV}^2$) and accompanied with the simulated
uncertainties reported by the LHeC study group \cite{Ref13}. The
H1 collaboration data with total errors for $Q^2=8.5$ and
$35~\mathrm{GeV}^2$ are shown in Fig.4.

In Fig. 5, we show a comparison between the longitudinal structure functions
in momentum space with the H1 Collaboration data at a fixed value
of the invariant mass $W$ (i.e. $W=230~\mathrm{GeV}$) at low
values of $x$. Figure 5 clearly demonstrates that the
Laplace transform method in momentum space provides correct
behaviors of the extracted longitudinal structure function  in
comparison with the LO and NLO analysis reported in \cite{Ref10}. As
can be seen in this figure, the results are comparable with the H1
data  and the NLO corrections to the Mellin transform method at
all $Q^{2}$ values.

Indeed, the momentum-space DGLAP evolution equations for structure functions measurable
in deeply inelastic scattering have some importance in contrast to the existing literature on
the subject where the evolution has been written in Mellin space.
In the momentum-space, there is no need to define a factorization scheme and also, the approach in terms of physical structure functions has the advantage of being more transparent in the parametrization of the initial conditions of the evolution.

\begin{figure}[h]
\includegraphics[width=0.6\textwidth]{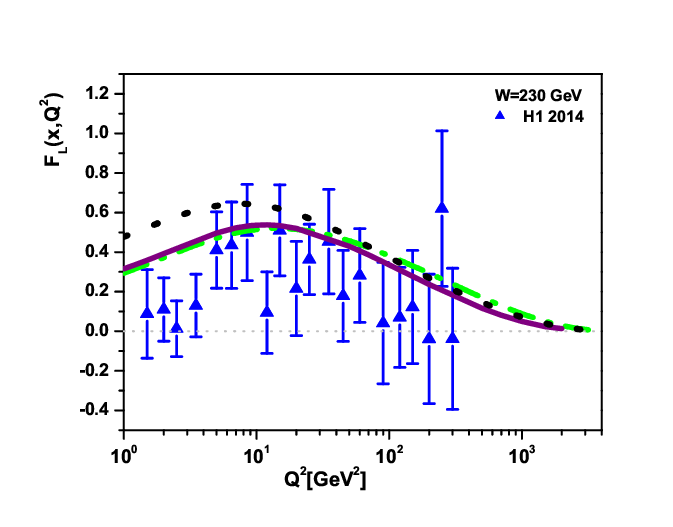}
\caption{The extracted longitudinal structure function (solid brown
curve) in momentum space at fixed value of the invariant mass $W$
($W=230~\mathrm{GeV}$) compared with the H1 Collaboration data
\cite{Ref42} as accompanied with total errors and the results in Refs. \cite{Ref8}
and \cite{Ref10} at the LO (dotted black curve) and NLO (dot-dashed green curve)
approximation. }\label{Fig4}
\end{figure}

\section{Conclusions}

We have presented a method based on the Laplace transform method to determine
the longitudinal structure function at the LO approximation in momentum space.
This method relies on the parametrization
of the function $F_{2}(x,Q^2)$ and its derivative $dF_{2}(x,Q^2)/d{\ln}Q^2$
within a kinematical region characterized by low values of the Bjorken variable $x$.
The $x$ dependence of $F_2(x,Q^2)$ and its evolution with $Q^2$ are determined
much better by the data than $F_L(x,Q^2)$, so this
method provides both a direct check on $F_L(x,Q^2)$ where measured,
and a way of extending $F_L(x,Q^2)$ into regions of $x$ and $Q^2$ where there are currently no data.
We find that the Laplace transform method in momentum space provides correct
behaviors of the extracted longitudinal structure function $F_{L}(x,Q^2)$  and that
our results for $F_{L}(x,Q^2)$ demonstrate comparability with data from the H1 Collaboration and
other results obtained using the Mellin transform method.

\section{ACKNOWLEDGMENTS}

 G.R.Boroun thanks M.Klein and N.Armesto for allowing access to
data related to simulated errors of the longitudinal structure
function at the Large Hadron Electron Collider (LHeC). Phuoc Ha would like to thank Professor
Loyal Durand for useful comments and invaluable support.


\section{Appendix}

The inverse Laplace transform

\begin{equation}
\widehat{f}(\upsilon) = {\mathcal{L}^{-1}}\bigg{[}f(s),s;\upsilon\bigg{]}=\frac{1}{2\pi i} \int_{- i\infty}^{i \infty} ds f(s) e^{s \upsilon}
\end{equation}
of the function
\begin{equation}
f(s) = \frac{(s+1)\psi(s+1)}{4+3s+s^2}
\end{equation}
can be evaluated analytically in terms of an infinite series, rapidly convergent except for $\upsilon$ near zero where it grows logarithmically.
The denominator in $f(s)$ has zeros at  $s_{\pm}=-\frac{3}{2} \pm i \frac{\sqrt{7}}{2}$. which lead to simple poles in $f(s)$  at those points.
The function $\psi(s+1)$ has simple poles with residue $-1$ at $s+1 = 0, -1, -2, ...$ \cite{Ref44} . There are no other sigularities in the integrand which decreases exponentially
rapidly for $s \rightarrow - \infty$ and {\cal Re}$(\upsilon) >0$. We can therefore close the integration contour in the left-half $s$ plane and evaluate the integral
as the sum of the residues at the poles multiplied by $2\pi i$ by Cauchy's residue theorem.

This gives
\begin{equation}
\widehat{f}(\upsilon) =
 \frac{s_+ +1}{s_+ - s_-}\psi(s_{+}+1) e^{s_+ \upsilon}
-  \frac{s_- +1}{s_+ - s_-}\psi(s_{-}+1) e^{s_- \upsilon} + \sum_{k=1}^{\infty}\frac{k}{(k+1)^2-3(k+1)+4}  e^{-(k+1) \upsilon} .
\end{equation}

Since
\begin{equation}
 \frac{s_+ +1}{s_+ - s_-}\psi(s_{+}+1) = -0.1704 +  1.211 i \, ,  \, \,  \, \, \,
 \frac{s_- +1}{s_+ - s_-}\psi(s_{-}+1)  = 0.1704 +  1.211 i  .
\end{equation}
we find
\begin{equation}
\widehat{f}(\upsilon) =- 2 e^{-\frac{3}{2}  \upsilon} \bigg{[} 0.1704 \cos  \big{(}\frac{\sqrt{7}}{2}\upsilon\big{)} +  1.211  \sin\big{(}\frac{\sqrt{7}}{2}\upsilon\big{)} \bigg{]}
+ \sum_{k=1}^{\infty}\frac{k}{(k+1)^2-3(k+1)+4}  e^{-(k+1) \upsilon} .
\end{equation}

For $\upsilon \ll 1$, $k$ large, the series is approximately
\begin{equation}
 \sum_{k}\frac{1}{k} e^{-(k+1) \upsilon} \approx e^{- \upsilon} \, {\rm ln} \, (1 -  e^{- \upsilon}) ,
\end{equation}
so diverges as ln $\upsilon$ for $\upsilon \to 0$.

\begin{table}[h]
\caption{ The effective parameters [7] at low $x$ for
$0.15~\mathrm{GeV}^{2}<Q^{2}<3000~\mathrm{GeV}^{2}$.}
\begin{tabular} {cccc}
\toprule \\  \multicolumn{2}{c}{parameters \quad \quad \quad ~~~~~~~~~~~~~~~~value}    \\ &&&\\ \hline \\ &&&\\
  $a_{0} $  &   \quad  $8.205\times 10^{-4}~~  \pm  4.62\times10^{-4} $  \\

  $a_{1} $  &   \quad   $-5.148\times 10^{-2}\pm 8.19\times10^{-3}$  \\

  $a_{2}$   &    \quad  $-4.725\times 10^{-3}\pm 1.01\times10^{-3}$   \\  &&&\\

 $b_{0}$   &   \quad   $2.217\times 10^{-3}\pm 1.42\times10^{-4} $ \\

 $b_{1}$   &   \quad   $1.244\times 10^{-2}\pm 8.56\times10^{-4}$  \\

 $b_{2}$    &    \quad  $5.958\times 10^{-4}\pm 2.32\times10^{-4} $ \\ &&& \\

$c_{1}$& \quad  $1.475\times 10^{-1}~\pm 3.025\times10^{-2}$ & &\\

$n$& \quad  $11.49\pm 0.99$ & &\\

$\lambda$& \quad  $2.430~\pm 0.153$ & &\\

$M^{2}$ & \quad $0.753 \pm 0.068~ \mathrm{GeV}^{2}$ & &\\

$\mu^2$ & \quad $ 2.82 \pm 0.290~ \mathrm{GeV}^{2}$ & &\\

$c_{0}$ & \quad $ 0.255 \pm 0.016$ & &\\

$\chi^{2}(\mathrm{goodness~ of~ fit})$ &  \quad  $0.95$ & &\\
\hline

\end{tabular}
\end{table}




\end{document}